\documentstyle[12pt]{article}

\newcommand{\be}{\begin{equation}}
\newcommand{\ee}{\end{equation}}
\newcommand{\bea}{\begin{eqnarray}}
\newcommand{\eea}{\end{eqnarray}}

\begin{document}
\title{
\begin{flushright}
{\small SMI-4-98 }
\end{flushright}
\vspace{2cm}
Non-extremal Localised  Branes
and \\ Vacuum Solutions in M-Theory }

\author{
I. Ya. Aref'eva${}^{\*}$, M. G. Ivanov${}^{\#}$,
O. A. Rytchkov${}^{\*}$\\ and \\ I. V. Volovich${}^{\*}$\\
\\${}^{\*}$ {\it  Steklov Mathematical Institute,}
\\ {\it Gubkin St.8, 117966 Moscow, Russia }
\\arefeva,rytchkov,volovich@mi.ras.ru\\
\\${}^{\#}$ { \it Moscow Institute of Physics and Technology}\\
            {\it Institutski per., 9, Dolgoprudnyi, Moscow reg.,
                Russia}    \\      mgi@mi.ras.ru
}
\date {$~$}
\maketitle
\begin {abstract}

Non-extremal overlapping p-brane  supergravity solutions localised in
their relative transverse coordinates are constructed.  The construction
uses an algebraic method of solving  the bosonic equations
of motion. It is shown that these non-extremal solutions can be
obtained from the extremal solutions by means of the superposition of
two deformation functions defined by vacuum solutions of M-theory.
Vacuum solutions of M-theory
including irrational powers of harmonic functions are discussed.

\end{abstract}
\newpage
\section{Introduction}
String theories and M-theory admit various solitonic extended-object
solutions \cite{To}-\cite{DR}.
The description of the p-brane solutions of supergravity
has been exploited to elucidate many nonperturbative aspects of string
theories as well as of gauge theories, see for example
\cite{Sch}-\cite{Kutasov}.  These extended solutions include
intersecting
extremal p-branes as well as non-extremal  or ``black'' p-branes
\cite{To}-\cite{DR}, \cite{Horowitz}-\cite{BC}.
A brane solution whose harmonic function is independent of a number
of transverse coordinates (relative transverse or overall transverse)
is said to be
delocalised over those directions. It was found that many delocalised
solutions have non-extremal generalisations
\cite{Horowitz}-\cite{Duff_Lu}, \cite{Duff_L_P}-\cite{BC}.

In this paper we are interested in more general localised solutions. It
will be shown that localised overlapping extremal solutions obtained in
\cite{Khuri,GKT} admit non-extremal generalisations.

Intersection rules \cite{PT,Ts} for M-branes and D-branes were found by
using the string theory representation of brane, duality and the
requirement of supersymmetry \cite{BREJS} or the no-force condition
\cite{Ts2}.  Another derivation of the intersection rules was obtained
in
\cite{AVVV,AR,Iv,Eng} by direct solving the bosonic equations of
motion
of the low-energy theory.  These intersection rules are universal, in
the
sense that they are not specific for some space-time dimension and
therefore do not require the supersymmetry.  Starting from these
intersection rules one can check \cite{AR,Eng} the harmonic
superposition rule and S,T-dualities.  These rules were first
obtained  for extremal branes and then were generalised
\cite{AIV,Ohta} to include the intersections of non-extremal branes.
Recently it was shown that the intersection rules have a simple
geometric meaning as the condition ensuring the symmetric space
property of the appropriate $\sigma$-model target space \cite{GalRy}.

In the last year there was an important progress in describing
non-perturbative phenomena in gauge theories using brane configurations.
In this approach branes are considered as configurations preserving a
part
of supersymmetry and one has to deal with intersecting configurations
having  a brane stretched between other branes. From the point
of view of application to non-perturbative study of
gauge theories \cite{HW}-\cite{Kutasov}, \cite{EGK}-\cite{EJS} via brane
consideration one has to find brane configurations as solutions of
equations of motion with special localisation properties, in
particular  solutions with one brane ending on another one.
It is a rather complicated problem to write explicitly gravity
solutions describing one brane ending on another brane \cite
{Strom,To2,Gaunt,Open}.  Brane configurations which in some
approximation satisfy
the desirable localisation properties
were considered in \cite{Khuri,GKT,Ts1,Gaunt} and further were
generalised
in \cite{Tatar}.
Also some examples of partially localised p-branes were constructed
in \cite{MI,LP}.

In this paper we present a construction  of non-extremal p-brane
solutions
which distinguishing characteristic is that branes are localised in the
relative transverse directions.
We derive an intersecting rule for pair-wise intersections of
non-extremal
branes.  As in the extremal case \cite{Tatar}, our non-extremal solutions
satisfy a characteristic equation
which is different from a standard characteristic equation of
intersecting
branes with harmonic functions depending only on overall transverse
directions \cite{AVVV}-\cite{Eng}.

Applying our general formulae presented in Section 2
to D=11 and D=10  we obtain in Section 3
non-extremal deformations of localised overlapping solutions, found
by Khuri \cite{Khuri}, Gauntlett, Kastor and
Traschen \cite{GKT}, and Tseytlin \cite{Ts1}.
It is
interesting that these deformations are specified by two different
vacuum
solutions.    Vacuum solutions
defining the deformation functions  satisfy to
a kind of harmonic superposition.
One deformation function corresponds to the
Schwarzshild-type vacuum solution and another to a new vacuum solution.
There are also more general deformations (see Section \ref{GSSec}  for
details).

\section{Localised Intersection of Two Non-extremal Branes}
Our construction
uses an algebraic method \cite{AVVV,AR} of solution of the bosonic
equations
of motion. It is convenient  to
start with the following expression for the low-energy bosonic
action in  D-dimensional space-time

\be
\label{act}
S=\frac{1}{2\kappa^2}\int\,d^Dx\sqrt{-g}(R-\frac12(\nabla\phi)^2-
\frac{e^{-\alpha_1\phi}}{2(d_1+1)!}F^2_{d_1+1}-
\frac{e^{-\alpha_2\phi}}{2(d_2+1)!}F^2_{d_2+1}),
\ee
where $F_{d_a+1}$ ($a=1,2$) is a $d_a$-form field strength,
$F=dA$, $\phi$ is a dilaton. The low-energy superstring
action contains also the Chern-Simons terms, but we omit them
since in the particular interesting examples that we will consider
in Section 3 they are irrelevant.

In order to find special solutions which
describe  localised p-branes we take  the line element
in the following form

$$
ds^2=e^{2(A_x+A_y)}(-f_xf_ydt^2+dz_1^2+\ldots+dz_{q-1}^2)+
e^{2(F_{1x}+F_{1y})}(f_x^{-1}dr_x^2+r_x^2d\Omega_{r_1-1})$$
\be
\label{metr0}
+e^{2(F_{2x}+F_{2y})}(f_y^{-1}dr_y^2+r_y^2d\Omega_{r_2-1})+
e^{2(B_x+B_y)}(du_1^2+\ldots+du_s^2).
\ee
Here  $z_i$, $i=1,...q-1$ belong to the intersection,
$x_{\gamma}$'s , $\gamma =1,...r_1$, $r_1\ge 3$ and $y_{\mu}$'s,
$\mu=1,...r_2$, $r_2\ge 3$ are the relative transverse coordinates and
the $u_k$'s, $k=1,...s$ are the overall transverse coordinates,
$d_1=q+r_1$, $d_2=q+r_2$.  We denote
$r_x=\sqrt{x^{\gamma}x^{\gamma}}$  and $r_y=\sqrt{y^{\mu}y^{\mu}}$.
All functions with the subscript $x$ depend on $r_x$ only, and the
functions with the subscript $y$ depend on $r_y$.

We will consider the intersection of two different localised p-branes.
They are coupled to $d_1$- and $d_2$-forms.
Consider for example, an electrically charged branes with the
following ans\"atze
for the forms  $$
F_{q+r_1+1}=h_1r_x^{r_1-1}\partial_{y}e^{D_{1x}+D_{1y}}dt\wedge
dz_1\wedge\cdots \wedge dz_{q-1}\wedge dr_x\wedge dr_y\wedge
vol_{r_1-1},$$
\be
F_{q+r_2+1}=h_2r_y^{r_2-1}\partial_{x}e^{D_{2x}+D_{2y}}dt\wedge
dz_1\wedge\cdots \wedge dz_{q-1}\wedge dr_x\wedge dr_y\wedge vol_{r_2-1},
\ee
where $vol_{r_1-1}$ and $vol_{r_2-1}$ are volume forms on the
$(r_1-1)$- and $(r_2-1)$-dimensional spheres,
$h_a$ are constants.
 $\partial_x$ and $\partial_y $ mean derivatives on $r_x$
and $r_y$, respectively.

Also we assume the following form for the dilaton
field \be \phi=\phi_x+\phi_y.  \ee Examining the Maxwell equations we
conclude (see Appendix) that \be e^{-D_{2x}}=H_x, \qquad
e^{-D_{1y}}=H_y, \ee where $H_x$ and $H_y$ are harmonic functions \be
H_x=1+\frac{Q_2}{r_x^{r_1-2}},\qquad H_y=1+\frac{Q_1}{r_y^{r_2-2}}.
\ee

The consistency of the Einstein equations gives (see Appendix )

\be
f_x=1-\frac{\mu_2}{r_x^{r_1-2}},\qquad
f_y=1-\frac{\mu_1}{r_y^{r_2-2}}.
\ee
This is in agreement with the vacuum solutions (see equation
(\ref{metrV0}) below).

The dilaton field is defined in terms of harmonic functions $H$
and $f$ via the following equations
\be
\label{phi'}
 \partial_x \phi_x=\frac{\alpha_2
Q_2(r_1-2)h_2^2}{2r_x^{r_1-1}f_x}H_x^{-1}+\frac{c_{\phi
1}}{r_x^{r_1-1}f_x},\qquad \partial_y \phi_y=\frac{\alpha_1
Q_1(r_2-2)h_1^2}{2r_y^{r_2-1}f_y}H_y^{-1}+\frac{c_{\phi
2}}{r_y^{r_2-1}f_y},
\ee
where $c_{\phi 1}$ and $c_{\phi 2}$ are some constants.

The metric functions $A_x$ and $A_y$
are defined by simple integration of the following equations

\be
\label{A'}
\partial_x
A_x=\frac{Q_2(r_1-2)t_2h_2^2}{r_x^{r_1-1}f_x}H_x^{-1}+\frac{c_{A1}}
{r_x^{r_1-1}f_x},\qquad
\partial_y A_y=\frac{Q_1(r_2-2)t_1h_1^2}{r_y^{r_2-1}f_y}H_y^{-1}+
\frac{c_{A2}}{r_y^{r_2-1}f_y},
\ee
where $c_{A1}$ and $c_{A2} $ are integration constants;
$t_1$,  $t_2$, $h_1$ and $h_2$ are constants  given by
\be
t_1=\frac{D-d_1-2}{2(D-2)},\qquad t_2=\frac{D-d_2-2}{2(D-2)}.
\ee

\be
h_1=\frac{2}{\sqrt{\Delta_1}}\sqrt{1+\frac{\mu_1}{Q_1}},\qquad
h_2=\frac{2}{\sqrt{\Delta_2}}\sqrt{1+\frac{\mu_2}{Q_2}},
\ee
where we use the usual notations
\be
\Delta_a=\alpha_a^2+\frac{2(D-d_a-2)d_a}{D-2}.
\ee

The metric functions  $F$ and $B$
are defined by  integration of the following equations

\be
\label{F1xF2y'}
\partial_x F_{1x}=-\frac{Q_2(r_1-2)u_2h_2^2}{r_x^{r_1-1}f_x}H_x^{-1}+
\frac{c_{F11}}{r_x^{r_1-1}f_x},\qquad
\partial_y F_{2y}=-\frac{Q_1(r_2-2)u_1h_1^2}{r_y^{r_2-1}f_y}H_y^{-1}+
\frac{c_{F22}}{r_y^{r_2-1}f_y},
\ee

\be
\label{F2xF1y'}
\partial_x
F_{2x}=\frac{Q_2(r_1-2)t_2h_2^2}{r_x^{r_1-1}f_x}H_x^{-1}+
\frac{c_{F21}}{r_x^{r_1-1}f_x},\qquad
\partial_y
F_{1y}=\frac{Q_1(r_2-2)t_1h_1^2}{r_y^{r_2-1}f_y}H_y^{-1}+
\frac{c_{F12}}{r_y^{r_2-1}f_y},
\ee

\be
\label{B'}
\partial_x B_x=-\frac{Q_2(r_1-2)u_2h_2^2}{r_x^{r_1-1}f_x}H_x^{-1}+
\frac{c_{B1}}{r_x^{r_1-1}f_x},\qquad
\partial_y B_y=-\frac{Q_1(r_2-2)u_1h_1^2}{r_y^{r_2-1}f_y}H_y^{-1}+
\frac{c_{B2}}{r_y^{r_2-1}f_y},
\ee

where  $c$'s are also integration
constants
and the constants $u_1$ and $u_2$ are given by
\be
u_1=\frac{d_1}{2(D-2)},\qquad u_2=\frac{d_2}{2(D-2)}.
\ee

The consistency of our ansatz with the equations of motion
imply the following restriction
on the parameters
\be
\label{INter}
\frac{\alpha_1\alpha_2}{2}+q+2-\frac{d_1d_2}{D-2}=0,
\ee
as well as  constraints on the integration constants

\be
\label{restr1'}
qc_{A1}+(r_1-2)c_{F11}+r_2c_{F21}+sc_{B1}=0,\qquad
qc_{A2}+r_1c_{F12}+(r_2-2)c_{F22}+sc_{B2}=0,
\ee

\be
\label{restr2'}
\alpha_2 c_{\phi1}+2qc_{A1}+2r_2c_{F21}+2(r_1-2)\mu_2=0,\qquad
\alpha_1 c_{\phi2}+2qc_{A2}+2r_1c_{F12}+2(r_2-2)\mu_1=0,
\ee

\be
\label{restr3'}
\alpha_1 c_{\phi1}+2qc_{A1}+2(r_1-2)c_{F11}+4c_{F21}=0,\qquad
\alpha_2 c_{\phi2}+2qc_{A2}+4c_{F12}+2(r_2-2)c_{F22}=0,
\ee

$$
\frac12 c_{\phi
1}^2+qc_{A1}^2+(r_1-2)c_{F11}^2+r_2c_{F21}^2+sc_{B1}^2+\mu_2(r_1-2)
(c_{A1}-c_{F11})=0,
$$
\be\label{restr4'}
\frac12 c_{\phi
2}^2+qc_{A2}^2+(r_2-2)c_{F22}^2+r_1c_{F12}^2+sc_{B2}^2+\mu_1(r_2-2)
(c_{A2}-c_{F22})=0,
\ee

$$
\mu_1\mu_2(\frac12r_1r_2-r_1-r_2+2)+
\mu_2(c_{A2}-c_{F12})(r_1-2)+\mu_1(c_{A1}
-c_{F21})(r_2-2)
$$
\be\label{restr5'}
+c_{\phi1}c_{\phi
2}+2qc_{A1}c_{A2}+2(r_1-2)c_{F11}c_{F12}+2(r_2-2)c_{F21}c_{F22}
+4c_{F12}c_{F21}+2sc_{B1}c_{B2}=0.
\ee

Now the main task is to solve the system of algebraic equations
(\ref{restr1'}), (\ref{restr2'}),  (\ref{restr3'}), (\ref{restr4'}),
(\ref{restr5'}) and find the appropriate constants. For simplicity
we will consider only particular cases.

\section{Localised Intersections in D=11 and D=10}
\subsection {Non-extremal localised M-branes}
\par
Equation (\ref{INter}) has the solution $\alpha_1=\alpha_2=0$, $q=2$,
$r_1=4$, $r_2=4$, $s=1$, which corresponds to intersecting M5-branes.
Our system of algebraic equations has the following solutions
$$
c_{\phi1}=0, \qquad c_{A1}=-\mu_2,\qquad c_{F11}=\mu_2,\qquad
c_{F21}=0,\qquad c_{B1}=0,
$$
\be
\label{sol1}
c_{\phi2}=0, \qquad c_{A2}=-\frac13\mu_1,\qquad
c_{F12}=-\frac13\mu_1,\qquad c_{F22}=\frac23\mu_1,\qquad
c_{B2}=\frac23\mu_1,
\ee
or
$$
c_{\phi1}=0, \qquad c_{A1}=-\frac13\mu_2,\qquad
c_{F11}=\frac23\mu_2,\qquad c_{F21}=-\frac13\mu_2,\qquad
c_{B1}=\frac23\mu_2,
$$
\be
\label{sol2}
c_{\phi2}=0, \qquad c_{A2}=-\mu_1,\qquad c_{F12}=0,\qquad
c_{F22}=\mu_1,\qquad c_{B2}=0.
\ee
Thus we get two metrics. The first solution
(\ref{sol1}) corresponds to the line element
$$ ds^2=H_x^{-1/3}
H_y^{-1/3} f_x^{-2/3}(-f_xf_ydt^2+dz^2)+ H_x^{2/3} H_y^{-1/3}
f_x^{1/3}(f_x^{-1}dr_x^2+r_x^2d\Omega_{3}) $$
\be
\label{ds2A}
+H_x^{-1/3}
H_y^{2/3} f_x^{1/3}(f_y^{-1}dr_y^2+r_y^2d\Omega'_{3}) +H_x^{2/3}
H_y^{2/3} f_x^{-2/3}du^2.
\ee
whereas
the second one  (\ref{sol2}) leads to the equivalent expression

$$
ds^2=H_x^{-1/3} H_y^{-1/3} f_y^{-2/3}(-f_xf_ydt^2+dz^2)+
H_x^{2/3} H_y^{-1/3} f_y^{1/3}(f_x^{-1}dr_x^2+r_x^2d\Omega_{3})
$$
\be
\label{ds2B}
+H_x^{-1/3} H_y^{2/3} f_y^{1/3}(f_y^{-1}dr_y^2+r_y^2d\Omega'_{3})
+H_x^{2/3} H_y^{2/3} f_y^{-2/3}du^2.
\ee

Solutions (\ref{ds2A}) and (\ref{ds2B}) describe a non-extremal
generalisation of the solution \cite{GKT,Gaunt} describing the
 M5-branes intersecting on a string.

We see that the harmonic functions are independent of the overall
transverse
direction $u$ and  depend only on the relative transverse directions.
As in the extremal case each
M5-brane is localised in the directions tangent to the other
M5-brane but is delocalised in the overall transverse direction that
separates them.

\subsection {Non-extremal localised NS-branes}

Equation (\ref{INter}) has also the following solution
$\alpha_1=1$, $\alpha_2=1$, $q=2$, $r_1=4$, $r_2=4$, $s=0$. It
describes the intersection of NS5-branes on a string.
In this case the integration constants are
$$
c_{\phi1}=0, \qquad c_{A1}=-\mu_2,\qquad c_{F11}=\mu_2,\qquad
c_{F21}=0,
$$
\be
\label{sol3}
c_{\phi2}=-\mu_1, \qquad c_{A2}=-\frac14\mu_1,\qquad
c_{F12}=-\frac14\mu_1,\qquad c_{F22}=\frac34\mu_1,
\ee
or
$$
c_{\phi1}=-\mu_2, \qquad c_{A1}=-\frac14\mu_2,\qquad
c_{F11}=\frac34\mu_2,\qquad c_{F21}=-\frac14\mu_2,
$$
\be
\label{sol4}
c_{\phi2}=0, \qquad c_{A2}=-\mu_1,\qquad c_{F12}=0,\qquad
c_{F22}=\mu_1.
\ee

We find the following solutions
$$ ds^2=H_x^{-1/4}
H_y^{-1/4} f_x^{-3/4}(-f_xf_ydt^2+dz^2)+ H_x^{3/4} H_y^{-1/4}
f_x^{1/4}(f_x^{-1}dr_x^2+r_x^2d\Omega_{3})
$$
\be
\label{ds2C}
+H_x^{-1/4}
H_y^{3/4} f_x^{1/4}(f_y^{-1}dr_y^2+r_y^2d\Omega'_{3}).
\ee

\be
e^{-\phi_x}=H_x^{1/2}f_x^{-1/2},\qquad e^{-\phi_y}=H_y^{1/2},
\ee
and

$$
ds^2=H_x^{-1/4} H_y^{-1/4} f_y^{-3/4}(-f_xf_ydt^2+dz^2)+
H_x^{3/4} H_y^{-1/4} f_y^{1/4}(f_x^{-1}dr_x^2+r_x^2d\Omega_{3})
$$
\be
\label{ds2D}
+H_x^{-1/4} H_y^{3/4} f_y^{1/4}(f_y^{-1}dr_y^2+r_y^2d\Omega'_{3}).
\ee
\be
e^{-\phi_x}=H_x^{1/2},\qquad e^{-\phi_y}=H_y^{1/2}f_y^{-1/2},
\ee

Solutions (\ref{ds2C}) and (\ref{ds2D})
give a non-extremal generalisation
of the solution \cite{Khuri} describing two NS5-branes
intersecting on a string.

\section{Vacuum Solutions in M-theory and Deformations
                \label{GSSec}}

A non-extremal deformation of an extremal solution is
performed by means of a vacuum solution.
Vacuum solution in M-theory is a Ricci flat metric
($R_{\mu\nu}=0$) in eleven dimensions, or the solution
with the cosmological constant
($R_{\mu\nu}=\Lambda g_{\mu\nu}$).

If one has a manifold $M^D$ which is the product of two
manifolds, $M^D=M^n\times M^{D-n}$, then we always have
a vacuum solution on $M^D$ as the sum of vacuum solutions
on $M^n$ and on $M^{D-n}$.
A simple example   is the
product of the standard Schwarzschild solution
and the Euclidean version of the
Schwarzschild solution
$$
ds_{V_{0}}^2=(-f_xdt^2+f_ydz_1^2+\ldots+dz_{q-1}^2)+
(f_x^{-1}dr_x^2+r_x^2d\Omega_{r_1-1})$$
\be
 \label{metrV0}
   +(f_y^{-1}dr_y^2+r_y^2d\Omega_{r_2-1})+
  (du_1^2+\ldots+du_s^2),
\ee
where $f_x$ and $f_y$ are $r_1$- and $r_2$-dimensional harmonic
functions.

Note that it is also possible to take the line element
defining a vacuum solution of the D-dimensional
gravity in another form $$
ds_{V}^2=f_x^{\nu_{0}}(-f_xf_ydt^2+dz_1^2+\ldots+dz_{q-1}^2)+
f_x^{\nu_{1}}(f_x^{-1}dr_x^2+r_x^2d\Omega_{r_1-1})$$
\be
\label{metrV00}
+f_x^{\nu_{2}}(f_y^{-1}dr_y^2+r_y^2d\Omega_{r_2-1})+
f_x^{\nu_{u}}(du_1^2+\ldots+du_s^2),
\ee
where the parameters $\nu_0$, $\nu_1$, $\nu_2$ and $\nu_u$ are
specified by an algebraic
system of non-linear equations
with the coefficients which depend on $D$, $r_1$ and $r_2$.
Generally, these solutions have naked singularities \cite{HGT}.

 In particular, for $D=11$ one of these solutions has the form
$$ ds^2= f_x^{-2/3}(-f_xf_ydt^2+dz^2)+
  f_x^{1/3}(f_x^{-1}dr_x^2+r_x^2d\Omega_{3}) $$
\be
 +f_x^{1/3}(f_y^{-1}dr_y^2+r_y^2d\Omega'_{3})
 +f_x^{-2/3}du^2.
\ee
This metric is nothing but the vacuum limit of the solution
(\ref{ds2A}).

 Notice also the following vacuum solutions of M-theory
(i.e. Ricci flat metrics in $D=11$ dimensions)
\bea
ds^2&=&
     {\rm f}_x^{-a}
       \left[
        -{\rm f}_y dt^2
        +{\rm f}_x^{-1}d\rho_x^2
        +\rho_x^2 d\Omega_2
        +{\rm f}_y^{-1}d\rho_y^2
        +\rho_y^2 d\Omega'_2
       \right]\\ \nonumber
   &+&{\rm f}_x^{4a+b+1} dz^2
    +{\rm f}_x^{-b}
       \left[ dv^2+ dw^2 \right]
    +{\rm f}_x^{a+b} du^2,
\eea
where
\bea
  {\rm f}_x=1-\frac{\mu_2}{\rho_x},~~~~~~~
  {\rm f}_y=1-\frac{\mu_1}{\rho_y},
\eea
and the real parameters $a$ and $b$ satisfy the equation
\be
\label{ab-eq} 11a^2+5a+2b^2+b+5ab=0.
\ee

In the general case the parameters $a$, $b$
are irrational. However there are  examples
when they are   rational

\bea
\label{set1}
   a=-\frac{1}{3},~~~~~~~
   b=-\frac{1}{3}~~~\mbox{or}~~~
   b= \frac{2}{3},\\
\label{set2}
   a=-\frac{1}{2},~~~~~~~
   b= \frac{1}{2}~~~\mbox{or}~~~
   b= \frac{1}{4}.
\eea
The pairs $a=-1/3$, $b=-1/3$ and
$a=-1/3$, $b=2/3$ give the same metric.

Other examples are
$a=-1/2$, $b=1/2$ and $a=-1/2$, $b=1/4$:
\bea
ds^2&=&
     {\rm f}_x^{1/2}
       \left[
        -{\rm f}_y dt^2
        +{\rm f}_x^{-1}d\rho_x^2
        +\rho_x^2 d\Omega_2
        +{\rm f}_y^{-1}d\rho_y^2
        +\rho_y^2 d\Omega'_2
       \right]\\ \nonumber
   &+&{\rm f}_x^{-1/2}
       \left[dz^2+dv^2+ dw^2 \right]
    +du^2,
\eea
\bea
ds^2&=&
     {\rm f}_x^{1/2}
       \left[
        -{\rm f}_y dt^2
        +{\rm f}_x^{-1}d\rho_x^2
        +\rho_x^2 d\Omega_2
        +{\rm f}_y^{-1}d\rho_y^2
        +\rho_y^2 d\Omega'_2
       \right]\\ \nonumber
   &+&{\rm f}_x^{-3/4} dz^2
    +{\rm f}_x^{-1/4}
       \left[ dv^2+ dw^2+du^2 \right].
\eea

These  vacuum solutions provide us with
the following  non-extremal solutions
of eleven dimensional supergravity
\bea
ds^2&=&{\rm H}_x^{-1/3}{\rm H}_y^{-1/3}
     \left\{
       -{\rm f}_x^{-a}{\rm f}_y dt^2
       +{\rm f}_x^{4a+b+1} dz^2
     \right\} \\ \nonumber
    &+&{\rm H}_x^{2/3}{\rm H}_y^{-1/3}
     \left\{
       {\rm f}_x^{-a}
       \left[
         {\rm f}_x^{-1}d\rho_x^2
        +\rho_x^2 d\Omega_2
       \right]
      +{\rm f}_x^{-b} dv^2
     \right\} \\ \nonumber
    &+&{\rm H}_x^{-1/3}{\rm H}_y^{2/3}
     \left\{
       {\rm f}_x^{-a}
       \left[
         {\rm f}_y^{-1}d\rho_y^2
        +\rho_y^2 d\Omega'_2
       \right]
      +{\rm f}_x^{-b} dw^2
     \right\}
    +{\rm H}_x^{2/3}{\rm H}_y^{2/3}
     {\rm f}_x^{a+b} du^2,
\eea
where
\bea
  {\rm H}_x=1+\frac{Q_2}{\rho_x},~~~~~~~
  {\rm H}_y=1+\frac{Q_1}{\rho_y}.
\eea

\section{Conclusion}
 In this paper we have presented a non-extremal
generalisation of localised overlapping brane solutions
in M-theory.
 The  harmonic functions specifying this solution are
independent of one  overall transverse direction  and depend on the
relative transverse directions.
 The M5-branes are
localised in the directions tangent to the other M5-brane but are
delocalised in the overall transverse direction that separates them.
 In the
extremal case there is a solution which contains an extra M2-brane which
overlaps each of the M5-branes.
 This solution still does not
describe a localised solution needed to the brane approach to
gauge theory, however it would be interesting to find it's non-extremal
deformation.

There exists also a new non-extremal solution which describes two
NS5-branes.
The  harmonic functions specifying this solution
 depend on the
relative transverse directions.  These NS5-branes are
localised inside the directions tangent to the other NS5-brane.

 We do not present here the solutions
which involve an additional M2-brane or NS1-brane.
 The method that we have
used in this paper can be generalised to these two cases and the
corresponding calculations will be a subject of a forthcoming paper.

 We also have presented here some more involved solutions,
which based on a more general class of vacuum solutions.
 These solutions  correspond to
smeared extremal brane configurations.
 Properties of these solutions require a further study.

 \newpage

$$~$$
{\bf ACKNOWLEDGMENTS}
$$~$$

I.A. and O.R.  are supported in part by  RFFI grant 96-01-00608 and
M.I. and I.V. are supported  in part by  RFFI grant 96-01-00312.

$$~$$

\appendix

{\Large\bf Appendix}
$$~$$

In Appendix we present the explicit derivation of the system
(\ref{restr1'}), (\ref{restr2'}),  (\ref{restr3'}), (\ref{restr4'}),
(\ref{restr5'}) presented in Section 2.

The equations of motion following from the action (\ref{act}) are
\be
\label{Einstein}
R_{MN}-\frac12
g_{MN}R=e^{-\alpha_1\phi}T_{MN}^{(F_{d_1+1})}+e^{-\alpha_2\phi}
T_{MN}^{(F_{d_2+1})}
+T_{MN}^{(\phi)}, \ee \be \label{Maxwell}
\partial_M\left(e^{-\alpha_1\phi}\sqrt{-g}F^{MM_1\ldots
M_{d_1}}\right)=0,\qquad
\partial_M\left(e^{-\alpha_2\phi}\sqrt{-g}F^{MM_1\ldots
M_{d_2}}\right)=0, \ee \be\label{dilaton}
\frac{1}{\sqrt{-g}}\partial_M(\sqrt{-g}g^{MN}\partial_M\phi)+
\frac{\alpha_1}{2(d_1+1)!}e^{-\alpha_1\phi}F^2_{d_1+1}+
\frac{\alpha_2}{2(d_2+1)!}e^{-\alpha_2\phi}F^2_{d_2+1}=0.
\ee
The energy-momentum tensors for the matter fields have the form
\be
T_{MN}^{(F_{d_a+1})}=\frac{1}{2d_a!}\left(F_{MM_1\ldots
M_{d_a}}{F_N}^{M_1\ldots M_{d_a}}-
\frac{g_{MN}}{2(d_a+1)}F^2_{d_a+1}\right), \ee \be
T_{MN}^{(\phi)}=\frac12\left(\partial_M\phi\partial_N\phi-\frac12g_{MN}
(\nabla\phi)^2\right).
\ee

 It is convenient to rewrite the Einstein equations in
 another form \be
R_{MN}=e^{-\alpha_1\phi}G_{MN}^{(F_{d_1+1})}+
e^{-\alpha_2\phi}G_{MN}^{(F_{d_2+1})}+G_{MN}^{(\phi)},
\ee
where
\be
G_{MN}^{(F_{d_a+1})}=T_{MN}^{(F_{d_a+1})}+
\frac{1}{2-D}g_{MN}T^{(F_{d_a+1})},
\qquad T^{(F_{d_a+1})}=g^{MN}T_{MN}^{(F_{d_a+1})}\ee
\be
G_{MN}^{(\phi)}=T_{MN}^{(\phi)}+\frac{1}{2-D}g_{MN}T^{(\phi)},
\qquad T^{(\phi)}=g^{MN}T_{MN}^{(\phi)}.\ee
For simplification we impose the gauge, that in the extremal case
corresponds to the Fock-De-Donder gauge
\be
\label{gauge}
qA_x+(r_1-2)F_{1x}+r_2F_{2x}+sB_x=0,\qquad
qA_y+r_1F_{1y}+(r_2-2)F_{2y}+sB_y=0.
\ee

For our ansatz (\ref{metr0}) the components of the Ricci tensor
in the chosen gauge (\ref{gauge}) are
$$
R_{tt}=-g_{tt}(e^{-2(F_{1x}+F_{1y})}\frac{1}{r_x^{r_1-1}}
\partial_x(r_x^{r_1-1}f_x\partial_x(A_x+\frac12\ln f_x))$$
\be
+e^{-2(F_{2x}+F_{2y})}\frac{1}{r_y^{r_2-1}}
\partial_y(r_y^{r_2-1}f_y\partial_y(A_y+\frac12\ln f_y))),
\ee
$$
R_{z_iz_i}=-g_{z_iz_i}(e^{-2(F_{1x}+F_{1y})}\frac{1}{r_x^{r_1-1}}
\partial_x(r_x^{r_1-1}f_x\partial_x A_x)$$
\be
+e^{-2(F_{2x}+F_{2y})}\frac{1}{r_y^{r_2-1}}
\partial_y(r_y^{r_2-1}f_y\partial_y A_y)),
\ee
$$
R_{u_ku_k}=-g_{u_ku_k}(e^{-2(F_{1x}+F_{1y})}\frac{1}{r_x^{r_1-1}}
\partial_x(r_x^{r_1-1}f_x\partial_x B_x)$$
\be
+e^{-2(F_{2x}+F_{2y})}\frac{1}{r_y^{r_2-1}}
\partial_y(r_y^{r_2-1}f_y\partial_y B_y)),
\ee

$$
R_{r_xr_x}=-\partial_x^2(\ln r_x^{r_1-1}f_x)-2(\partial_x F_{1x}-
\frac12\partial_x\ln f_x)^2+\partial_x (2F_{1x}-\ln f_x)
\partial_x
(2F_{1x}+\ln r_x^{r_1-1})
$$ $$
-(\partial_x A_x+\frac12\partial_x \ln f_x)^2
-(q-1)(\partial_x A_x)-
(\partial_x F_{1x}-\frac12 \partial_x\ln f_x)^2
-(r_1-1)(\partial_x F_{1x}+\partial_x \ln r_x)^2
$$
\be
-r_2(\partial_x F_{2x})^2-s(\partial_x B_x)^2-
\partial_x^2(F_{1x}+
\frac12\ln f_x)-\partial_x(F_{1x}+\frac12\ln
f_x)\partial_x(\ln r_x^{r_1-1}f_x),
\ee

$$
R_{r_yr_y}=-\partial_y^2(\ln r_y^{r_2-1}f_y)-2(\partial_y F_{2y}-
\frac12\partial_y\ln f_y)^2+\partial_y (2F_{2y}-\ln f_y)\partial_y
(2F_{2y}+\ln r_y^{r_2-1})
$$ $$
-(\partial_y A_y+\frac12\partial_y \ln f_y)^2
-(q-1)(\partial_y A_y)-(\partial_y F_{2y}-\frac12 
\partial_y\ln f_y)^2
-(r_2-1)(\partial_y F_{2y}+\partial_y \ln r_y)^2
$$
\be
-r_1(\partial_y F_{1y})^2-s(\partial_y B_y)^2-\partial_y^2(F_{2y}+
\frac12\ln f_y)-\partial_y(F_{2y}+\frac12\ln
f_y)\partial_y(\ln r_y^{r_2-1}f_y),
\ee

$$
R_{r_xr_y}=-2\partial_x F_{2x}\partial_y F_{1y}+
\partial_x F_{2x}\partial_y (2F_{2y}+\ln r_y^{r_2-1})
$$ $$ +
\partial_y F_{1y}\partial_x(2F_{1x}+\ln r_x^{r_1-1})
-\partial_x(A_x+\frac12\ln f_x)\partial_y(A_y+\frac12\ln f_y)$$
$$
-(q-1)\partial_x A_x\partial_y A_y-\partial_x(F_{1x}-\frac12 f_x)
\partial_y F_{1y}-(r_1-1)\partial_x(F_{1x}+\ln r_x)\partial_y F_{1y}
$$
\be
-\partial_x F_{2x}\partial_y(F_{2y}-\frac12\ln f_y)-(r_2-1)\partial_x
F_{2x}\partial_y(F_{2y}+\ln r_y)-s\partial_x B_x\partial_y B_y,
\ee
where $\partial_x$ and $\partial_y $ as in Section 2 mean
the derivatives on $r_x$
and $r_y$, respectively.

Now let us analyse the components $G_{MN}$. For our ansatz
$G_{MN}^{(\phi)}$ has three non-trivial components
\be
G_{r_xr_x}^{(\phi)}=\frac12(\partial_x\phi_x)^2,\qquad
G_{r_xr_y}^{(\phi)}=\frac12\partial_x\phi_x\partial_y\phi_y,\qquad
G_{r_yr_y}^{(\phi)}=\frac12(\partial_y\phi_y)^2.
\ee
Also we have
$$
e^{-\alpha_1\phi}G_{MN}^{(F_{d_1+1})}+
e^{-\alpha_2\phi}G_{MN}^{(F_{d_2+1})}=
$$ \be
-g_{MN}(
t_1h_1^2e^{-2(F_{2x}+F_{2y})}S_xH_y^2(\partial_y e^{D_{1y}})^2+
t_2h_2^2e^{-2(F_{1x}+F_{1y})}S_yH_x^2(\partial_x e^{D_{2x}})^2)
\ee
for $tt$-, $zz$-, $r_xr_x$- and  $r_yr_y$-components,
$$
e^{-\alpha_1\phi}G_{MN}^{(F_{d_1+1})}+
e^{-\alpha_2\phi}G_{MN}^{(F_{d_2+1})}$$
\be
=g_{MN}(
u_1h_1^2e^{-2(F_{2x}+F_{2y})}S_xH_y^2(\partial_y e^{D_{1y}})^2+
u_2h_2^2e^{-2(F_{1x}+F_{1y})}S_yH_x^2(\partial_x e^{D_{2x}})^2)
\ee
for $uu$-components,
$$
e^{-\alpha_1\phi}G_{MN}^{(F_{d_1+1})}+
e^{-\alpha_2\phi}G_{MN}^{(F_{d_2+1})}$$ \be=-g_{MN}(
t_1h_1^2e^{-2(F_{2x}+F_{2y})}S_xH_y^2(\partial_y e^{D_{1y}})^2-
u_2h_2^2e^{-2(F_{1x}+F_{1y})}S_yH_x^2(\partial_x e^{D_{2x}})^2)
\ee
for the components corresponding to $r_1-1$-dimensional block and
$$
e^{-\alpha_1\phi}G_{MN}^{(F_{d_1+1})}+
e^{-\alpha_2\phi}G_{MN}^{(F_{d_2+1})}=$$ \be -g_{MN}(
-u_1h_1^2e^{-2(F_{2x}+F_{2y})}S_xH_y^2(\partial_y e^{D_{1y}})^2+
t_2h_2^2e^{-2(F_{1x}+F_{1y})}S_yH_x^2(\partial_x e^{D_{2x}})^2)
\ee
for the components corresponding to $r_2-1$-dimensional block.
We use the following notations
\be
\label{H}
H_x^2=\exp(-\alpha_2\phi_x-2qA_x-2r_2F_{2x}),\qquad
H_y^2=\exp(-\alpha_1\phi_y-2qA_y-2r_1F_{1y}),
\ee
\be
S_x=\exp(-\alpha_1\phi_x-2qA_x-2r_1F_{1x}+2D_{1x}),\qquad
S_y=\exp(-\alpha_2\phi_y-2qA_y-2r_2F_{2y}+2D_{2y}).
\ee

The Maxwell equations for our field configuration read
\be
\label{M_self}
\partial_x(H_x^2r_x^{r_1-1}\partial_x e^{D_{2x}})=0,\qquad
\partial_y (H_y^2 r_y^{r_2-1}\partial_y e^{D_{1y}})=0,
\ee
\be
\label{M_cross}
\partial_x(S_x e^{2F_{1x}-2F_{2x}-D_{1x}}\partial_ye^{D_{1y}})=0,\qquad
\partial_y(S_y e^{2F_{2y}-2F_{1y}-D_{2y}}\partial_xe^{D_{2x}})=0.
\ee

To solve the equations (\ref{M_self}) we set
\be
\label{D2xD1y}
e^{D_{2x}}=H_x^{-1},\qquad e^{D_{1y}}=H_y^{-1}.
\ee
It means that the functions should be harmonic
\be
H_x=1+\frac{Q_2}{r_x^{r_1-2}},\qquad H_y=1+\frac{Q_1}{r_y^{r_2-2}}.
\ee

Considering the dilaton equation as well as the Einstein equations
it is not difficult to observe that the variables
$x$ and $y$ could be easily separated if
\be
\label{sep}
S_x=1,\qquad S_y=1.
\ee
\be
\label{rest1}
2D_{1x}=\alpha_1\phi_x+2qA_x+2r_1F_{1x},\qquad
2D_{2y}=\alpha_2\phi_y+2qA_y+2r_2F_{2y}.
\ee

It leads to further simplifications of the Maxwell equations
(\ref{M_cross}) from  which we now get

\be
\label{rest2}
D_{1x}=2F_{1x}-2F_{2x},\qquad
D_{2y}=2F_{2y}-2F_{1y}.
\ee

We also assume that the parts of the dilaton equation  containing
differentiation on $r_x$ and $r_y$ are equalised separately
$$
\frac{1}{r_x^{r_1-1}}\partial_x(r_x^{r_1-1}f_x\partial_x\phi_x)=
\frac{\alpha_2}{2}h_2^2H_x^{2}(\partial_x e^{D_{2x}})^2,$$
\be\label{dilaton2}
\frac{1}{r_y^{r_2-1}}\partial_y(r_y^{r_2-1}f_y\partial_y\phi_y)=
\frac{\alpha_1}{2}h_1^2H_y^{2}(\partial_y e^{D_{1y}})^2.
\ee

Using (\ref{D2xD1y}) one can integrate both sides of  these  equations
to get
\be
\label{phi}
 \partial_x \phi_x=\frac{\alpha_2
Q_2(r_1-2)h_2^2}{2r_x^{r_1-1}f_x}H_x^{-1}+\frac{c_{\phi
1}}{r_x^{r_1-1}f_x},\qquad \partial_y \phi_y=\frac{\alpha_1
Q_1(r_2-2)h_1^2}{2r_y^{r_2-1}f_y}H_y^{-1}+\frac{c_{\phi
2}}{r_y^{r_2-1}f_y},
\ee
where $c_{\phi 1}$ and $c_{\phi 2}$ are some constants.
Equations (\ref{rest1}) and (\ref{rest2}) can be used to
express  $F_{1x}$ and  $F_{2y}$ through the other
variables
\be\label{F}
2(r_1-2)F_{1x}=-\alpha_1\phi_x-2qA_x-4F_{2x},\qquad
2(r_2-2)F_{2y}=-\alpha_2\phi_y-2qA_y-4F_{1y}.
\ee

Now let us consider the Einstein equations. The equation on the
$tt$-component under assumption (\ref{sep}) and after
a  separate equalising the terms depending
on $r_x$ and $r_y$  gives
$$
\frac{1}{r_x^{r_1-1}}\partial_x(r_x^{r_1-1}f_x
\partial_x(A_x+\frac12\ln f_x))=
t_2h_2^2H_x^{2}(\partial_x e^{D_{2x}})^2,$$
\be
\label{tt}
\frac{1}{r_y^{r_2-1}}\partial_y(r_y^{r_2-1}f_y\partial_y(A_y+
\frac12\ln f_y))=
t_1h_1^2H_y^{2}(\partial_y e^{D_{1y}})^2,
\ee
where $t_1$ and $t_2$ are constants
\be
t_1=\frac{D-d_1-2}{2(D-2)},\qquad t_2=\frac{D-d_2-2}{2(D-2)}.
\ee
The
equations on the $zz$-components under the same assumptions are
\be
\label{zz}
\frac{1}{r_x^{r_1-1}}\partial_x(r_x^{r_1-1}f_x\partial_x A_x)=
t_2h_2^2H_x^{2}(\partial_x e^{D_{2x}})^2,\qquad
\frac{1}{r_y^{r_2-1}}\partial_y(r_y^{r_2-1}f_y\partial_y
A_y)= t_1h_1^2H_y^{2}(\partial_y e^{D_{1y}})^2.
\ee
The consistency condition of the equations (\ref{tt}) and (\ref{zz})
gives the equations on the functions $f_x$ and $f_y$
\be
\frac{1}{r_x^{r_1-1}}\partial_x(r_x^{r_1-1}f_x\partial_x(\ln
f_x))=0,\qquad
\frac{1}{r_y^{r_2-1}}\partial_y(r_y^{r_2-1}f_y\partial_y(\ln f_y))=0.
\ee
Thus we have
\be
f_x=1-\frac{\mu_2}{r_x^{r_1-2}},\qquad
f_y=1-\frac{\mu_1}{r_y^{r_2-2}}.  \ee The substitution of
(\ref{D2xD1y}) in (\ref{zz}) gives
\be\label{A} \partial_x
A_x=\frac{Q_2(r_1-2)t_2h_2^2}{r_x^{r_1-1}f_x}H_x^{-1}+\frac{c_{A1}}
{r_x^{r_1-1}f_x},\qquad
\partial_y A_y=\frac{Q_1(r_2-2)t_1h_1^2}{r_y^{r_2-1}f_y}H_y^{-1}+
\frac{c_{A2}}{r_y^{r_2-1}f_y},
\ee
where $c_{A1}$ and $c_{A2} $ are integration constants.

The Einstein equations on the $uu$-components  yield
$$
\frac{1}{r_x^{r_1-1}}\partial_x(r_x^{r_1-1}f_x\partial_x B_x)=
-u_2h_2^2H_x^{2}(\partial_x e^{D_{2x}})^2,$$
\be\label{uu}
\frac{1}{r_y^{r_2-1}}\partial_y(r_y^{r_2-1}f_y\partial_y
B_y)=-u_1h_1^2H_y^{2}(\partial_y e^{D_{1y}})^2,  \ee
or
\be\label{B}
\partial_x B_x=-\frac{Q_2(r_1-2)u_2h_2^2}{r_x^{r_1-1}f_x}H_x^{-1}+
\frac{c_{B1}}{r_x^{r_1-1}f_x},\qquad
\partial_y B_y=-\frac{Q_1(r_2-2)u_1h_1^2}{r_y^{r_2-1}f_y}H_y^{-1}+
\frac{c_{B2}}{r_y^{r_2-1}f_y},
\ee
where  $c_{B1}$ and $c_{B2} $ are some  arbitrary constants
and the constants $u_1$ and $u_2$ are given by
\be
u_1=\frac{d_1}{2(D-2)},\qquad u_2=\frac{d_2}{2(D-2)}.
\ee

The consideration of the $\varphi\varphi$-components of the
Einstein equations where $\varphi$ is one of the angles on the
$(r_1-1)$- or $(r_2-1)$-sphere except the relations considered above
gives

\be
\label{F2x}
\frac{1}{r_x^{r_1-1}}\partial_x(r_x^{r_1-1}f_x\partial_x F_{2x})=
t_2h_2^2H_x^{2}(\partial_x e^{D_{2x}})^2,\qquad
\frac{1}{r_y^{r_2-1}}\partial_y(r_y^{r_2-1}f_y\partial_y
F_{1y})=t_1h_1^2H_y^{2}(\partial_y e^{D_{1y}})^2,  \ee
or
\be\label{F2xF1y}
\partial_x
F_{2x}=\frac{Q_2(r_1-2)t_2h_2^2}{r_x^{r_1-1}f_x}H_x^{-1}+
\frac{c_{F21}}{r_x^{r_1-1}f_x},\qquad
\partial_y
F_{1y}=\frac{Q_1(r_2-2)t_1h_1^2}{r_y^{r_2-1}f_y}H_y^{-1}+
\frac{c_{F12}}{r_y^{r_2-1}f_y},
\ee

$$
\frac{1}{r_x^{r_1-1}}\partial_x(r_x^{r_1-1}f_x\partial_x F_{1x})=
-u_2h_2^2H_x^{2}(\partial_x e^{D_{2x}})^2,$$
\be
\frac{1}{r_y^{r_2-1}}\partial_y(r_y^{r_2-1}f_y\partial_y
F_{2y})=-u_1h_1^2H_y^{2}(\partial_y e^{D_{1y}})^2,  \ee
or
\be\label{F1xF2y}
\partial_x F_{1x}=-\frac{Q_2(r_1-2)u_2h_2^2}{r_x^{r_1-1}f_x}H_x^{-1}+
\frac{c_{F11}}{r_x^{r_1-1}f_x},\qquad
\partial_y F_{2y}=-\frac{Q_1(r_2-2)u_1h_1^2}{r_y^{r_2-1}f_y}H_y^{-1}+
\frac{c_{F22}}{r_y^{r_2-1}f_y}.
\ee

Now let us examine the consistency conditions of equations
(\ref{phi}),  (\ref{A}), (\ref{B}), (\ref{F2xF1y}), (\ref{F1xF2y})
on one hand and algebraic restrictions (\ref{gauge}),
(\ref{H}) and (\ref{F}) on the other. The gauge condition (\ref{gauge})
gives the relations between the integration constants
\be
\label{restr1}
qc_{A1}+(r_1-2)c_{F11}+r_2c_{F21}+sc_{B1}=0,\qquad
qc_{A2}+r_1c_{F12}+(r_2-2)c_{F22}+sc_{B2}=0.
\ee From the equation (\ref{H}) we have
\be
\label{restr2}
\alpha_2 c_{\phi1}+2qc_{A1}+2r_2c_{F21}+2(r_1-2)\mu_2=0,\qquad
\alpha_1 c_{\phi2}+2qc_{A2}+2r_1c_{F12}+2(r_2-2)\mu_1=0.
\ee
\be
h_1=\frac{2}{\sqrt{\Delta_1}}\sqrt{1+\frac{\mu_1}{Q_1}},\qquad
h_2=\frac{2}{\sqrt{\Delta_2}}\sqrt{1+\frac{\mu_2}{Q_2}},
\ee
where we use the usual notations
\be
\Delta_a=\alpha_a^2+\frac{2(D-d_a-2)d_a}{D-2}.
\ee
The third type of relations (\ref{F}) gives us the restriction
on the parameters
\be
\label{inter}
\frac{\alpha_1\alpha_2}{2}+q+2-\frac{d_1d_2}{D-2}=0,
\ee
as well as new constraints on the integration constants
\be
\label{restr3}
\alpha_1 c_{\phi1}+2qc_{A1}+2(r_1-2)c_{F11}+4c_{F21}=0,\qquad
\alpha_2 c_{\phi2}+2qc_{A2}+4c_{F12}+2(r_2-2)c_{F22}=0.
\ee

Let us consider the equations on the $r_xr_x$- and
$r_yr_y$-components. After separation of the variables we have
$$
\frac12(\partial_x\phi_x)^2+q(\partial_xA_{x})^2+(r_1-2)
(\partial_xF_{1x})^2+r_2(\partial_xF_{2x})^2
+s(\partial_xB_{x})^2$$
\be
-\partial_x\ln
f_x(\partial_xF_{1x}-\partial_xA_{x})=\frac12h_2^2f_x^{-1}H_x^2
(\partial_x e^{D_{2x}})^2,
\ee
$$
\frac12(\partial_y\phi_y)^2+q(\partial_yA_{y})^2+(r_2-2)
(\partial_yF_{2y})^2+r_1(\partial_yF_{1y})^2
+s(\partial_yB_{y})^2$$
\be
-\partial_y\ln
f_y(\partial_yF_{1y}-\partial_yA_{y})=\frac12h_1^2f_y^{-1}H_y^2
(\partial_y e^{D_{1y}})^2,
\ee
These equations give us another restrictions

$$
\frac12 c_{\phi
1}^2+qc_{A1}^2+(r_1-2)c_{F11}^2+r_2c_{F21}^2+sc_{B1}^2+\mu_2(r_1-2)
(c_{A1}-c_{F11})=0,
$$
\be\label{restr4}
\frac12 c_{\phi
2}^2+qc_{A2}^2+(r_2-2)c_{F22}^2+r_1c_{F12}^2+sc_{B2}^2+\mu_1(r_2-2)
(c_{A2}-c_{F22})=0,
\ee
Besides, we obtain the restriction on the constants $c_{\phi1}$ and
$c_{\phi2}$, particularly, if $\alpha_1\ne\alpha_2$ we have to
put $c_{\phi1}=c_{\phi2}=0$.

Finally we have to consider the $r_xr_y$-component of the
Einstein equations $$
\frac12\partial_x\phi_{x}\partial_y\phi_{y}+
q\partial_xA_{x}\partial_yA_{y}
+2\partial_xF_{2x}\partial_yF_{1y}+
(r_1-2)\partial_xF_{1x}\partial_y
F_{1y}+(r_2-2)\partial_xF_{2x}\partial_yF_{2y}+
s\partial_xB_{x}\partial_yB_{y}
$$
\be
\label{r_xr_y}
-\frac12\partial_x\ln
f_x(\partial_yF_{1y}-\partial_yA_{y})-\frac12\partial_y\ln
f_y(\partial_xF_{2x}-\partial_xA_{x})+\frac14\partial_x\ln
f_x\partial_y\ln
f_y=0.
\ee
Taking into account the equations (\ref{phi}), (\ref{A}), (\ref{B}),
(\ref{F1xF2y}) and
(\ref{F2xF1y}) after straightforward calculations we obtain that
the equation (\ref{r_xr_y}) is satisfied if the constants
are subjects of the additional relation
$$
\mu_1\mu_2(\frac12r_1r_2-r_1-r_2+2)+
\mu_2(c_{A2}-c_{F12})(r_1-2)+\mu_1(c_{A1}
-c_{F21})(r_2-2)
$$
\be\label{restr5}
+c_{\phi1}c_{\phi
2}+2qc_{A1}c_{A2}+2(r_1-2)c_{F11}c_{F12}+2(r_2-2)c_{F21}c_{F22}
+4c_{F12}c_{F21}+2sc_{B1}c_{B2}=0.
\ee

\end{document}